\documentclass[12pt, letterpaper]{article}

\usepackage{graphicx}
\usepackage[utf8]{inputenc}
\usepackage{algorithm, algorithmicx, algpseudocode}
\usepackage{multirow}
\usepackage{enumitem}
\usepackage{url}

\title{Detection of Malicious Android Applications:
	Classical Machine Learning vs. Deep Neural Network
	Integrated with Clustering}
\author{
Hemant Rathore, Sanjay K. Sahay, Shivin Thukral\\
Department of CS\&IS, BITS Pilani, Goa, India\\
\texttt{\{hemantr, ssahay, h20180057\}@goa.bits-pilani.ac.in}
\and
Mohit Sewak\\
Security \& Compliance Research, Microsoft R\&D, India\\
\texttt{mohit.sewak@microsoft.com}
}

\date{} 

\begin{document}
\maketitle  

% \author{Hemant Rathore \and Sanjay K. Sahay \and Shivin Thukral \and Mohit Sewak}
%
% \authorrunning{Rathore et al.}
% First names are abbreviated in the running head.
% If there are more than two authors, 'et al.' is used.
%
% \institute{BITS Pilani, Department of CS \& IS, Goa Campus, India\\
% \email{\{hemantr, ssahay, f201500350, p20150023\}\\@goa.bits-pilani.ac.in}}

% \maketitle              % typeset the header of the contribution
%
\begin{abstract}
Today anti-malware community is facing challenges due to ever-increasing sophistication and volume of malware attacks developed by adversaries. Traditional malware detection mechanisms are not able to cope-up against next-generation malware attacks. Therefore in this paper, we propose effective and efficient Android malware detection models based on machine learning and deep learning integrated with clustering.  We performed a comprehensive study of different feature reduction, classification and clustering algorithms over various performance metrics to construct the Android malware detection models. Our experimental results show that malware detection models developed using Random Forest eclipsed deep neural network and other classifiers on the majority of performance metrics. The baseline Random Forest model without any feature reduction achieved the highest AUC of $99.4\%$. Also, the segregating of vector space using clustering integrated with Random Forest further boosted the AUC to $99.6\%$ in one cluster and direct detection of Android malware in another cluster, thus reducing the curse of dimensionality. Additionally, we found that feature reduction in detection models does improve the model efficiency (training and testing time) many folds without much penalty on effectiveness of detection model.

% \keywords{Android Malware \and Cyber Security \and Deep Neural Network \and Machine Learning \and Malware Detection \and Static Analysis.}
\end{abstract}

\section{Introduction}\label{Introduction}

Mobile phone and internet are increasingly becoming an integral part of our daily life. A recent report suggests that there are around $5.13$ billion mobile phone users, which is more than $65\%$ of the world’s population \cite{globaldr}. The mobile phone market is currently dominated by the Android Operating System (OS), which runs on more than $70\%$ of the devices \cite{idc}. Surprisingly there are $8.485$ billion mobile connections which are even more than the world`s current population \cite{smartphones}. Also, around $50\%$ of internet traffic flows through mobile phones \cite{globaldr}. Mobile phones hold a large amount of personal user data like documents, pictures, contacts, messages etc. and are an easy target of malware designers to steal the above information.

Malware (\textbf{Mal}icious Soft\textbf{ware}) is a universal name given to any program performing any malicious activity. The first known malware (\textit{Creeper}, $1971$) was designed for TENEX OS to display taunting messages \cite{ye2017survey}. In the initial years, attackers would design malware to show their knowledge or for fun. These attacks would pose low-security threat risk to the system, but today it is a profit-driven industry. Individuals, groups, states, etc. perform malware attacks motivated by gains associated with it. The first Android OS malware (\textit{ANDROIDOS\_DROIDSMS.A}\footnote{\url{https://www.trendmicro.com/vinfo/us/threat-encyclopedia/malware/ANDROIDOS\_DROIDSMS.A}}, $2010$) was developed to perform SMS fraud \cite{ye2017survey}. According to G DATA CyberDefense, $214,327$ new Android malware were detected in $2012$ \cite{gdata}. Since then, there has been an exponential growth in the velocity of incoming Android malware. Another recent report by G DATA CyberDefense shows $4$ million new malware was detected in $2018$ \cite{gdata}. Last year Symantec blocked $10,573$ malicious Android applications every day which were trying to steal user’s information \cite{symantec}.

Anti-malware detection engines developed by Avast, Kaspersky, McAfee, Symantec etc. are the primary defence to protect the genuine user from malware attacks \cite{ye2017survey}. Traditionally, these engines were based on signature based malware detection mechanism. A signature is a short unique byte sequence used to identify a particular malware \cite{ye2017survey}. However, the signature generation is often human-driven and time-consuming process. Also, malware can easily evade signature based detection by modifying a small amount of malicious code without affecting the overall semantic \cite{rastogi2013droidchameleon}. Automated malware development toolkits viz. \textit{Zeus}\footnote{\url{https://usa.kaspersky.com/resource-center/threats/zeus-virus}} can generate thousands of variants of malware in a single day using obfuscation techniques \cite{ye2017survey}. Creating different signatures for the detection of all the malware variants is an infeasible task. Also, signature based detection does not provide security against a zero-day attack\footnote{\url{https://www.avast.com/c-zero-day}}. Thus, Heuristic-based detection was developed where domain experts write generic rule/pattern to discriminate malicious and benign applications \cite{griffin2009automatic}. Ideally, rules/pattern should be comprehensive so that they do not increase false positives and false negatives which is often unacceptable for any real-world deployment and achieving it is a tough task. Thus scientists and researchers are developing new intelligent Android malware detection system based on machine learning and deep learning techniques which are less human-driven, and more effective, efficient, and scalable.

Malware analysis and detection using machine learning and deep learning is a two-step process: feature engineering followed by classification/clustering. The performance of any malware detection system is highly dependent on the feature set, and the category of the machine learning/deep learning algorithm used to build the model. Researchers have used feature set like permission \cite{sun2016sigpid} \cite{arp2014drebin}, intent \cite{wu2012droidmat} \cite{buczak2015survey}, API calls \cite{hou2016droiddelver} \cite{xu2016toward}, opcode \cite{sewak2020doom} in the past for detecting malicious Android applications. After feature vector creation, classification/clustering algorithm is used to build the malware detection model. The selection of the feature set is critical since they might contribute to many limitations of the malware detection systems. For example, a detection model based on Android permission can detect malicious activity performed through permission module only. Often researchers combine multiple feature sets and create huge vector space to build the detection model. The disadvantages of these models are that they suffer from the curse of dimensionality, need of a sophisticated classifier to construct the detection model, a large model size for any real-time deployment, huge train time and test time. Also, explainability of these models is very poor. Therefore in this paper, we propose to use Android opcodes as a feature for malware detection. Opcodes are at the lowest abstraction of Android OS architecture\footnote{\url{https://developer.android.com/guide/platform}} and can detect malware attack designed at higher OS abstraction as well. We performed an extensive feature vector analysis with different feature reduction techniques like attribute sub-selection and new attribute creation methods. Also, we used three categories of classification algorithms viz. classical machine learning, ensemble learning and deep neural network to build malware detection models and analyzed them using different performance metrics. Lastly, to further improve performance, we used clustering to divide the feature vector into smaller spaces (clusters) and then built classification models on each one of them to reduce the curse of dimensionality. Therefore in this paper, we performed extensive work for effective and efficient Android malware detection models and made the following contributions:

\begin{itemize}
	\item[$\bullet$] Our baseline Android malware detection model developed using random forest achieved the highest AUC ($99.4$). The reduced feature model based on random forest with only $12\%$ opcodes obtained an AUC of $99.1$ and reduction in the training and testing time by $47\%$ and $8\%$ respectively.
	\item[$\bullet$] Segregating of vector space using clustering followed by classification further boosted the performance of malware detection models with a higher AUC of $99.6$ (using random forest) in one cluster and direct detection of Android malware in another cluster. Experimental results show that the construction of a single complex malware detection model on the complete dataset may overfit/underfit the data. Thus segregation of vector space (clusters) integrated with classification can reduce the above effect by reducing the curse of dimensionality.
	\item[$\bullet$] Android malware detection models based on autoencoder and deep neural network perform poorly as compared to machine learning models. They achieved comparable effectiveness in detection but with substantially higher train and test time. 
\end{itemize}

The rest of the paper is organized as follows. Related work based on malware analysis and detection is discussed in Section 2. The proposed experimental setup and various performance metrics used in the paper are introduced in section 3. Experimental results including feature reduction, classification models and segregation of the feature vector (using clustering) integrated with classification are explained in Section 4. Finally, Section 5 concludes the paper and discuss future work.

\section{Related Work}\label{Related Work}

Malware analysis and detection is an endless competition between malware designers and the anti-malware community. Traditional malware detection systems are based on the signature, heuristic, and cloud-based engines which are not able to cope-up with new-age sophisticated malware attacks. Thus anti-malware community is trying to construct next-generation Android malware detection systems based on machine learning and deep learning \cite{ye2017survey} \cite{sewak2020overview} \cite{ganesh2017cnn}. These systems are developed using the two-step process (1) Feature Engineering (2) Classification. Thus we have also divided the literature survey into two lines of research \textbf{(1) Feature Engineering} consists of feature extraction and feature selection. Hou et al. \cite{hou2016droiddelver} developed DroidDelver (2016) for Android malware detection based on API calls as features on comodo dataset. Wang et al. \cite{wang2016droiddeeplearner} proposed DroidDeepLearner, which used features like permission and API call extracted from Android applications. DroidMat \cite{wu2012droidmat} extracted features including permission, application component, intent, and API call for building the detection model. Drebin \cite{arp2014drebin} extracted permission, intent, app component, API call and network address from the Android applications for the construction of malware detection. Most of the above work combines multiple features set for the development of Android malware detection models. For example, Drebin used roughly $545,000$ different features extracted from Android applications to build the malware detection model and thus suffers heavily from the curse of dimensionality. Literature shows that various feature selection methods viz. Document Frequency \cite{ye2017survey}\cite{henchiri2006feature}, Information Gain \cite{ye2017survey}\cite{peng2005feature}, Hierarchical Feature Selection \cite{ye2017survey}, Max-Relevance Algorithm \cite{ye2017survey}\cite{henchiri2006feature}, etc. have been used for construction of malware detection system but the cost of feature reduction is rarely discussed. On the other hand, Lindorfer et al. in Andrubis \cite{lindorfer2014andrubis} gathered more than $1$ million Android applications dated between $2010$ to $2014$. They found exponential growth in the number of Android applications (both malicious and benign) in the same time frame. They also found that the average number of permission requested by benign applications is less than malicious ones. Sarma \cite{sarma2012android} also found that certain permissions like \textit{READ\_SMS}, \textit{WRITE\_CONTACTS} etc. are used extensively by malicious applications as compare to benign application. Yang et al. \cite{yang2015appcontext} tracked the flow of intent to differentiate malicious and benign behaviour. Sharma et al. \cite{sharma2018investigation} performed grouping of Android applications using permissions and then performed classification using opcodes in each group. Zhou et al. \cite{zhou2012dissecting} found $86\%$ repacked benign applications to be containing some malicious payload and $93\&$ of malicious application exhibited bot-like capabilities. They also show that the anti-virus engine performed poorly and are unable to detection next-generation Android malware.   \textbf{(2) Building classification model:} After the development of feature vector, various machine learning and deep learning algorithms can be used for the construction of the Android malware detection system. Arp et al. \cite{arp2014drebin} in Drebin used $545,000$ different features and support vector machine as the classification algorithm for the construction of the Android malware detection model, which achieved 93.80\% detection accuracy. DroidMat \cite{wu2012droidmat} used permission, intent and API call as features combined with k-means, naive bayes, k-nearest neighbour algorithms to propose a detection model which achieved the f-measure of 91.8\%. Sharma et al. \cite{sharma2018investigation} on Drebin dataset first performed permission-based grouping and then used opcode for construction of detection model. They used a variety of tree-based classification algorithm like J48, functional trees, NBTree, logistic model tree, random forest and attained $79.27\%$ accuracy with functional trees. Sewak et al. \cite{sewak2018investigation} on Malicia dataset used opcode as the feature and explored models based on random forest and deep neural network to achieve $99.78\%$ detection accuracy with the random forest model. DroidDeepLearner \cite{wang2016droiddeeplearner} developed a deep belief network for malware detection and achieved 92.67\%  accuracy in malware classification. Li et al. \cite{li2018android} on Drebin dataset used kirin rules and attained $94.29\%$ recall in Android malware detection. DroidDelver \cite{hou2016droiddelver} used decision tree, naive bayes, support vector machine and deep learning to build detection models and attained 94.04\% of detection accuracy with decision tree-based model.

\section {Experimental Setup and Performance Metrics}\label{experiment_setup}

In this section, we will discuss the dataset (malicious and benign apps), feature vector generation and various evaluation metrics used to design Android malware detection models.

\subsection {Malicious and Benign apps}

We downloaded real Android malware applications from the \textit{Drebin} project \cite{arp2014drebin}. Arp et al. downloaded $131,611$ Android applications from Google Play Store\footnote{\url{https://play.google.com/store?hl=en}} and various other sources for the construction of the Drebin dataset. All these downloaded applications were inspected by a list of $10$ popular antivirus scanners and then labelled as malware or benign. Additionally, the dataset also includes all the $1,260$ malicious applications from the Android Malware Genome Project \cite{zhou2012dissecting}. The final published Drebin dataset consists of $5,560$ malicious Android applications from more than $50$ different malware families.

For this project, we collected $9,823$ Android apps from the Google Play Store between August 2018 to December 2018. We used the services of VirusTotal\footnote{\url{https://www.virustotal.com/}} to label applications as malware or benign. VirusTotal is a subsidiary of Chronicle/Google which provides APIs to scan Android applications by a list of antiviruses (viz. AVG, McAfee, Symantec, etc.). We scanned all the downloaded Android applications with VirusTotal and labelled them as benign or malicious. A sample was labelled benign if all the antiviruses from virustotal.com platform declared it as benign. We deleted applications which are reported malicious by VirusTotal. Finally, we marked a set of $5,592$ Android applications as benign. Thus the final dataset used in the experiments contains 5,560 malicious and 5,592 benign apps.

\subsection {Generating Feature Vector} \label{Generating Feature Vector}

The feature set is the backbone of any machine learning/deep learning based Android malware detection system. Features extraction can be performed using static or dynamic analysis of Android apps \cite{yan2018survey} \cite{egele2008survey}. From the malware designer perspective, malicious activities can be performed from any abstracted layer in the Android OS architecture\footnote{\url{https://developer.android.com/guide/platform}}. However, Android malware detection model based on permission feature set can detect malicious activity performed through permission only. Thus feature set generated from a particular higher Android OS abstraction is bound to have limitations. On the other hand, any malicious activity designed on any abstracted level will have to execute on a set of opcodes to complete its desired effect. Thus we have used opcode frequency generated by static analysis as the feature vector to design Android malware detection system.

Android applications can be dissembled using Apktool\footnote{\url{https://ibotpeaches.github.io/Apktool/}} to generate its assembly code. The disassembled code consists of AndroidManifest.xml file, smali files, classes.dex files, images, and other application components. A \textit{master opcode list} was generated by using a parser which lists the $256$ Dalvik opcodes (represented in hexadecimal, ranging from $00$ to $FF$) in each Android application. For example, hexadecimal values of opcodes \textit{nop} and \textit{move} are $00$ and $01$ respectively. The above list was also verified from the official Android website\footnote{\url{https://developer.android.com/reference/dalvik/bytecode/Opcodes}}. Finally, a different parser scanned through each of the Android applications and generated its opcode frequency vector. The final feature vector for the complete dataset is $11138$ $\times$ $256$ where a row denotes an Android app, and a column denotes the frequency of a particular opcode.

\subsection {Evaluation Metrics}

We have employed both unsupervised and supervised learning methods for building the Android malware detection system. Following are different evaluation metrics used to build detection models.

\subsubsection{Methods/Metrics to find the optimal number of clusters:}\label{cluster_metric} Clustering is an unsupervised learning method used to understand the pattern/structure of the data. In most clustering algorithm, the number of clusters should be provided by the programmer and finding it in the dataset is a very subjective and challenging task. Domain knowledge, metrics, visualization tools can be used to decide the optimal number of clusters, but there is no single thumb rule for it. We have used the Silhouette Score and Calinski-Harabasz Index to find an optimal number of clusters in clustering algorithms.

\begin{itemize}
\item[$\bullet$]\textbf{Sum of Square Error} is used to find the optimal number of clusters (k) in the k-means clustering algorithm. In this method, we calculated the Sum of Square Error (SSE) as a function of k which can be used to understand the compactness of the cluster(s). The value of k at an elbow-like bend in the plot can be considered as its optimal value. The mathematical representation of SSE is as follows:

\begin{eqnarray} \label{sse}
SSE &=& \sum_{i=1}^{k} \sum_{x\in{c_i}} dist^2(x, c_i)
\end{eqnarray}
where $k$ is number of clusters, $c_i$ is the centroid of cluster $i$ and $x$ is all the points in cluster $i$

\item[$\bullet$]\textbf{Silhouette Score} is used to measure the similarity of a point within the cluster (cohesion) compared to other clusters (separation). It can vary from $+1$ to $-1$ where a higher positive value indicates the given point is well inside the cluster boundary, zero value indicates that the point is on the cluster boundary, and a negative value indicates it is dissimilar to the assigned cluster. We have used Euclidean distance as the distance metric to measure the silhouette score.

\begin{eqnarray}
	s(i) = \frac{b(i)-a(i)}{max\{a(i),b(i)\}}
\end{eqnarray}

where $a(i)$ is the average distance between point $i$ and all the points in the cluster (cohesion) and $b(i)$ is the average distance between point $i$ and all the points in the neighbouring cluster(s) (separation)

\item[$\bullet$]\textbf{Calinski-Harabasz Index} is the ratio of between-cluster covariance and within-cluster covariance.

\begin{eqnarray}
	CH-Index = \frac{SSE_B}{SSE_W} \times \frac{N-k}{k-1}
\end{eqnarray}

where $SSE_W$ is within-cluster covariance calculated by eq(\ref{sse}), $SSE_B$ is the sum of $SSE$ of all clusters minus $SSE_W$, $k$ is the number of clusters, and $N$ is the total number of data points.
\end{itemize}

\subsubsection{Classification Algorithm Performance Metrics}\label{classification_metric}

To evaluate the performance of the Android malware detection models, we have used following metrics that are derived from the confusion matrix.

\begin{itemize}
\item[$\bullet$]\textbf{Accuracy} is the ratio of correct predictions (malware and benign) by the total predictions.
%\\$Accuracy = \left(TP + TN \right) / \left(TP + FP + TN + FN\right) $

\begin{eqnarray}
	Accuracy &=& \frac{TP + TN}{TP + FP + TN + FN}
\end{eqnarray}

\item[$\bullet$]\textbf{Recall (TPR)} measures how many actual malignant applications are classified as malware by the detection model.

\begin{eqnarray}
	TPR &=& \frac{TP}{TP + FN}
\end{eqnarray}

\item[$\bullet$]\textbf{Specificity (TNR)} measures how many actual benign applications are classified as benign by the detection model. 
\begin{eqnarray}
	TNR &=& \frac{TN}{TN + FP}
\end{eqnarray}

\item[$\bullet$]\textbf{F-measure (F1)} is harmonic mean of recall and precision.

\begin{eqnarray}
	F1 &=& 2 \times {\frac{recall \times precision}{recall + precision}} 	
\end{eqnarray}

\item[$\bullet$]\textbf{Area Under The Curve (AUC)} illustrates the ability of malware detection model to classify benign application as benign and malware application at malware. A Perfect classification model has an ideal AUC value of $1$.
\end{itemize}

\section{Experimental Analysis and Results}

\begin{figure*}[htbp]
    \centering
	\includegraphics[width=\linewidth]{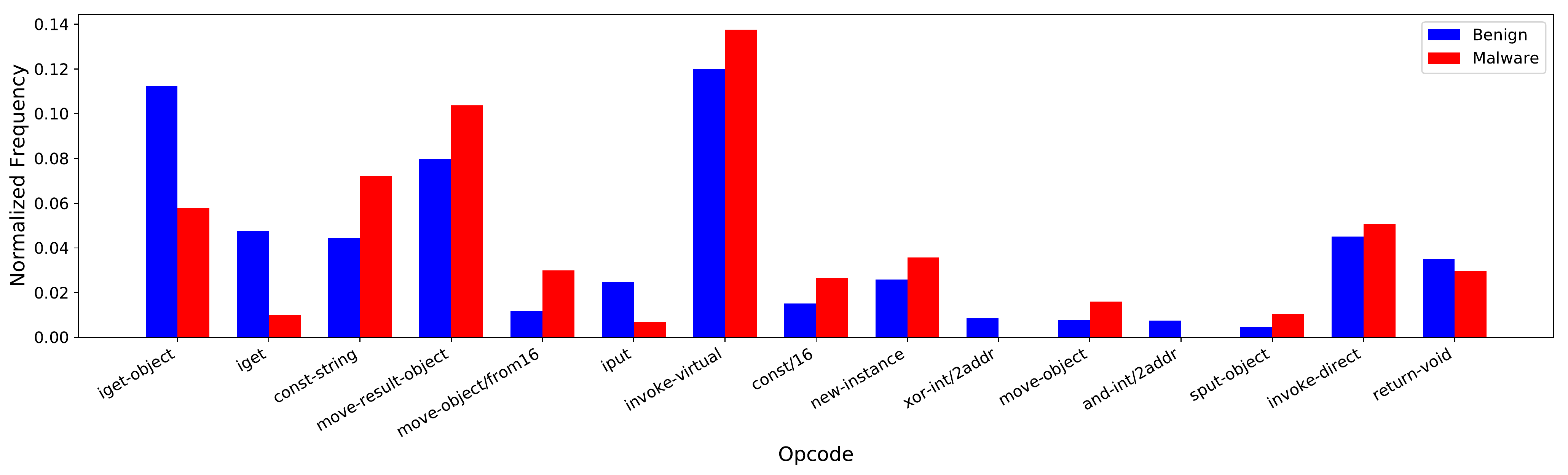}
	\caption{Top $15$ opcodes having maximum normalized frequency difference}
	\label{fig/normalize_freq}
\end{figure*}

Firstly the opcode frequencies extracted from Android applications are embedded in the feature vector space. Analyzing the feature vector, we found that $44$ opcodes out of $256$ have never been used by any Android application (malicious or benign). Out of the above set, $26$ opcodes are named as \textit{unused\_$\ast\ast$} and are reserved for future use by Android OS. Further analyzing the feature vector, we found that all the Android applications (malware and benign) are not of the same size. Thus the number of opcodes in each application will also not be the same. We propose the following algorithm $1$ for normalizing the feature vector. The algorithm first normalizes the opcodes based on Android application size and then based on their category (benign or malware). It also returns the k-prominent opcodes for further analysis to design Android malware detection models. Figure \ref{fig/normalize_freq} shows top $15$ prominent opcodes after preprocessing and sorted in descending order.  Opcode \textit{iget-object} has a maximum normalized frequency difference in malware and benign applications followed by \textit{iget}, \textit{const-string} and so on. It also signifies that the distribution of opcodes in malware and benign files are not the same. Some opcodes may be dominant in malicious Android applications, while others are more frequent in benign applications.

\begin{algorithm}[htbp]
    \caption{Pseudocode for generating normalized features vector}
    \begin{flushleft}
	\textbf{Input}: Pre-processed data \\
	$\mathbf{N_B}$: Number of benign applications \\
	$\mathbf{N_M}$: Number of malware applications \\
	$\mathbf{k}$: Total number of prominent opcodes required \\
	$\mathbf{n}$: Total number of opcodes in dataset\\
	\textbf{Output} : List of sorted prominent opcodes
	\begin{algorithmic}
		\For {each file f}
		\State Compute sum of opcodes $\mathbf{Op_j}$ in file $\mathbf{f_i}$ and normalize it
		\State ${f_i ( Op_j ) = f_i ( Op_j ) / ( \sum_{j=1}^{n} f_i (Op_j))}$
		\EndFor
		
		\For {each opcode $\mathbf{Op_j}$}
		\State Compute sum of frequencies of $\mathbf{Op_j}$ across all benign files $\mathbf{f_i}$ and normalize it to $\mathbf{F_B}(Op_j)$
		\State ${F_B ( Op_j ) = ( \sum_{i=1}^{N_B} f_i (Op_j)) / N_B}$

		\State Compute sum of frequencies of $\mathbf{Op_j}$ across all malware files $\mathbf{f_i}$ and normalize it to $\mathbf{F_M}(Op_j)$
		\State ${F_M ( Op_j ) = ( \sum_{i=1}^{N_M} f_i (Op_j)) / N_M}$
		\EndFor
		
		\For {all opcodes $\mathbf{Op_j}$}
		\State Find the difference of the normalized frequencies for each opcode $\mathbf{D(Op_j)}$
		\State ${D ( Op_j ) = \mid F_B (Op_j) - F_M (Op_j) \mid}$
		\EndFor
		\State return $k$ number of prominent opcodes with high $D(Op)$
		
	\end{algorithmic}
	\end{flushleft}
\end{algorithm}

% \begin{figure*}[ht]
% 	\centerline{\includegraphics[width=0.8\linewidth]{fig/normalize_freq}}
% 	\caption{Top $15$ opcodes having maximum normalized frequency difference}
% 	\label{fig/normalize_freq}
% \end{figure*}

Correlation is a statistical measure to indicate the relationship between two or more variables. Correlation analysis of Android malware dataset suggests that few opcode sets tend to occur more frequently in malicious apps. The top three opcode pairs in malicious apps having maximum correlations are \{\textit{monitor-enter} \& \textit{monitor-exit}\}, \{\textit{add-double} \& \textit{sub-double}\} and \{\textit{new-instance} \& \textit{invoke-direct}\} with correlation values as $0.9993$, $0.9846$ and $0.9830$ respectively. Similarly, co-correlation analysis of benign and malware apps separately indicates that the intersection of opcode pairs having high correlation values in both malicious and benign applications is negligible

% %
% \begin{figure}[ht]
% 	\centering
% 	\includegraphics[width=1\linewidth]{fig/top_corr_mal_opcodes}
% 	\caption{Top correlations in opcodes of malicious applications}
% 	\label{fig/top_corr_mal_opcodes}
% \end{figure}
% %

\subsection{Feature Reduction} \label{Feature Reduction}
Since the feature vector consists of $256$ opcodes with attribute type as continuous, thus the Android malware detection models built without any feature reduction is most likely to suffer from the curse of dimensionality. We used both categories of feature reduction methods: attribute sub-selection (viz. variance threshold) and new feature creation (viz. principal component analysis and autoencoders).

\begin{figure}[htbp]
	\centering
	\includegraphics[width=0.6\linewidth]{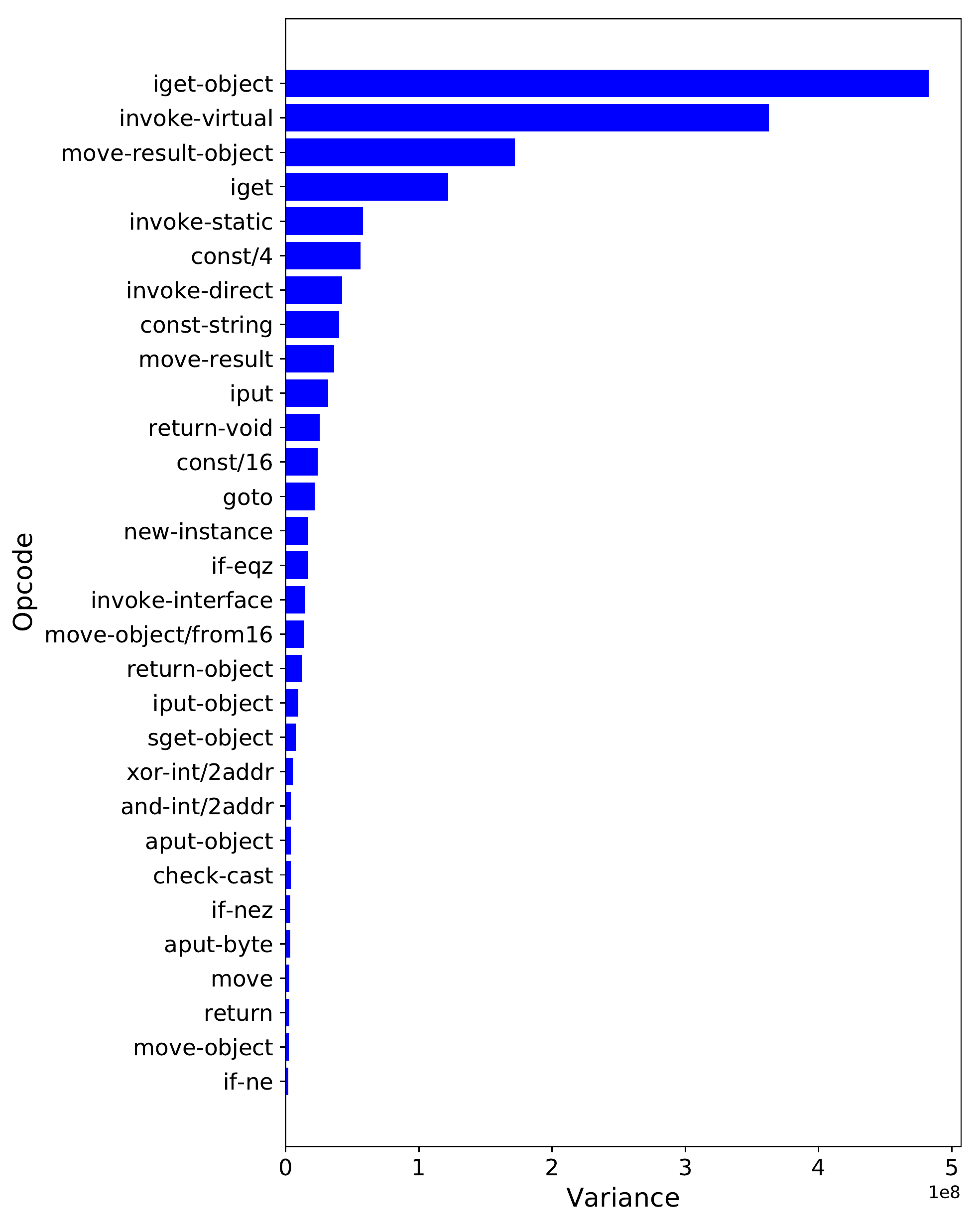}
	\caption{Top $30$ opcodes having maximum variance}
	\label{fig/VT_30}
\end{figure}

\textbf{Variance Threshold (VT)} is an unsupervised method used to remove noise and less relevant attributes from the dataset, thus removing opcodes having less prediction power. Figure \ref{fig/VT_30} shows the top $30$ opcodes having maximum variance in the complete dataset. Opcode \textit{iget-object} has maximum variance followed by \textit{invoke-virtual} and \textit{move-result-object}. We removed opcodes having less prediction power, and thus the reduced feature vector from VT consists of top $30$ opcodes having maximum variance.

% %
% \begin{figure}[ht]
% 	\centering
% 	\includegraphics[width=1\linewidth]{fig/VT}
% 	\caption{Variance in opcode frequencies}
% 	\label{fig/VT}
% \end{figure}

%

\begin{figure}[htbp]
	\centering
	\includegraphics[width=0.7\linewidth]{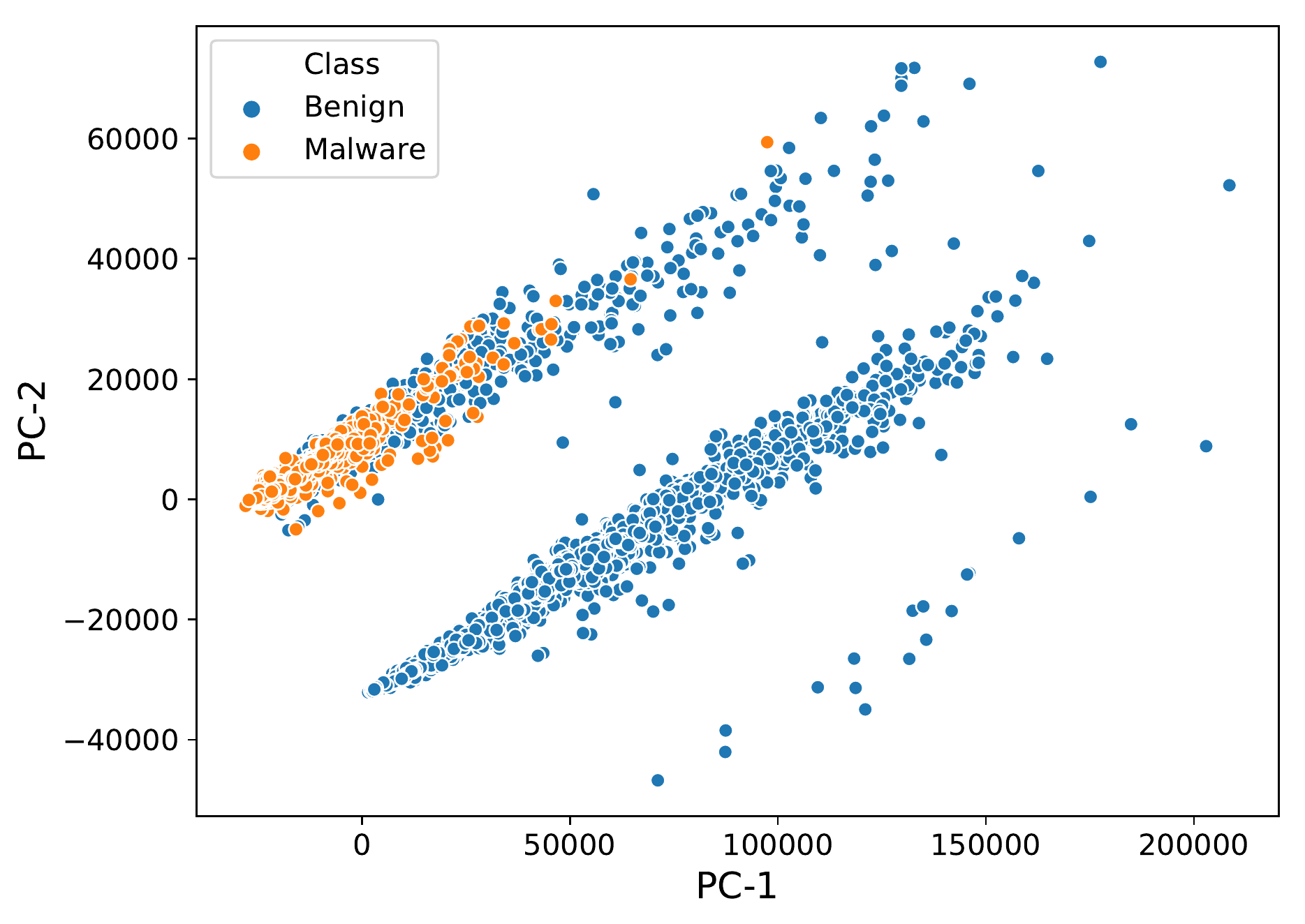}
	\caption{PC-1 and PC-2 from PCA}
	\label{fig/pca_dis}
\end{figure}

\textbf{Principal Component Analysis (PCA)} is a data transformation technique that maps an input to a set of new orthogonal vectors. We performed PCA analysis on the original data and found that the number of principal components as $15$ to be stable with different detection models. The figure \ref{fig/pca_dis} shows the two principal compeonts having the highest variance represented in $x$ and $y$ axis of the plot. The figure shows that most of the data points belonging to Android malware are restricted to one single region in the plot, while benign data points are distributed across the ranges of both the axes. Thus we can infer that malicious applications from a particular malware family will contain similar opcodes with limited frequency ranges. In contrast, benign applications will be spread more exhaustively in the vector space.

% %
% \begin{figure}[ht]
% 	\centering
% 	\includegraphics[width=1\linewidth]{fig/pca}
% 	\caption{Principal Component Analysis for feature reduction}
% 	\label{fig/pca}
% \end{figure}
% %

%
\textbf{Auto Encoder} is an unsupervised learning technique based on deep neural networks used for dimensionality reduction. Autoencoder receives the data in the input layer and compresses it with encoding layers until the bottleneck layer. Further decoding layers in the network will uncompress the data to match the original data closely. Since the feature set consists of $256$ attributes, the input and output layer in all the autoencoders contains $256$ nodes. We have designed two different auto-encoders to produce two different data transformations on feature set:

\begin{itemize}
	
	\item[$\bullet$] \textbf{1-layer Auto Encoder (AE-1L)} consists of the input layer, one encoding layer, and the output/decoding layer. The overall architecture of the AE-1L is 256-64-256
	
	\item[$\bullet$] \textbf{3-layer (stacked) Auto Encoder (AE-3)} consists of the input layer, three encoding layers of sizes $64$, $32$, $16$ nodes consecutively, and three decoding layers of sizes $32$, $64$ and $256$ nodes. The overall architecture of AE-3L is 256-64-32-16-32-64-256
	
\end{itemize}

All encoding and decoding layers of the two autoencoders are fully-connected and use the exponential linear unit (ELU) function for activation. ELU is used over the rectified linear unit (ReLU) because it converges faster due to the smooth function. Since the autoencoders are prone to overfitting, dropout is used for better generalization. A drop probability of $0.4$ was set to all hidden layers of AE-3 (layers with 32 and 64 nodes) except the bottleneck layer. All the autoencoders use Adam Optimizer with a learning rate of $0.001$ and are trained over $100$ epochs with a batch size of $32$ samples. A validation set of $20$\% of the entire samples was separated to verify the results. As the goal is to reconstruct the input, Mean Squared Error (MSE) function was used to compare the loss between the original input and the produced output. During the AE training, the training and validation loss decreases steeply in the initial epochs since the autoencoder can perform reconstruction with a fast pace, after which further decrease in loss is prolonged because the bottleneck layer does not allow $100$\% reconstruction.

% %
% \begin{figure}[ht]
% 	\centering
% 	\includegraphics[width=1\linewidth]{fig/AE-1L_train_mse}
% 	\caption{Sample plot for AE-1L showing training and validation loss}
% 	\label{fig/AE-1L_train_mse}
% \end{figure}
% %

\subsection{Baseline Malware Detection Models} \label{Malware Detection}

We used three diverse categories of classification algorithms to build the Android malware detection system. The first set consists of classical machine learning classifiers viz. Decision Tree (DT), k-Nearest Neighbors (kNN) and Support Vector Machine (SVM) while the second set consists of ensemble classifier viz. Random Forest (RF), and Adaptive Boosting (AdaBoost). The third set uses deep learning to build three models based on deep neural network.

\textbf{Malware Detection using Classical Machine Learning:}
DT is a tree-based classification algorithm used for predictive modelling. We used the GINI criterion for splitting the tree nodes with minimum samples required for the split set as two. The DT models were built without any restriction on tree depth or minimum applications needed in the leaf node. For the construction of kNN models, we considered five nearest neighbours and Euclidean distance was used to identify the neighbours. Also, equal weights were assigned to all the neighbours for voting. We train the SVM models using a linear kernel with penalty term $C$ and tolerance for stopping criterion set to $1.0$ and $1e-3$ respectively. Table \ref{big_table} and section \ref{discussion} discuss the performance of the Android malware detection models built using the above classification algorithms with different feature reduction techniques based on different evaluation metrics.

\textbf{Malware Detection using Ensemble Learning:}
RF is an ensemble learning method which employs bagged DT. Our RF model uses an ensemble of $100$ DTs, with the GINI index used as a criterion for determining best split. There was no constraint imposed on tree depth, the maximum number of leaf nodes in DT or number of samples required for a node split. Our AdaBoost model was an ensemble of $100$ estimators, where the base estimator is a DT classifier with maximum depth as $1$. Also, the learning rate from each model was set to $1$. SAMME.R was used as the boosting algorithm. Performance of the ensemble learning models for Android malware detection are shown and discussed in Table \ref{big_table} and section \ref{discussion} respectively.

\textbf{Malware Detection using Deep Neural Network:}
For classification of malicious Android applications, we train three different deep neural networks models with varying (shallow to deep network) architecture:
 
 \begin{itemize}
 	\item[$\bullet$] \textbf{2-layer Deep Neural Network (DNN-2L)} contains one hidden layers with $64$ nodes. The final architecture of DNN-2L is 256-64-1
 	\item[$\bullet$] \textbf{4-layer Deep Neural Network (DNN-4L)} is a deeper network containing three hidden layers with $128$, $32$ and $8$ nodes respectively. The final architecture of DNN-4L is 256-128-32-8-1
 	\item[$\bullet$] \textbf{7-layer Deep Neural Network (DNN-7L)} contains six hidden layers with $128$, $64$, $32$, $8$, and $4$ nodes. The final architecture of DNN-7L is 256-128-64-32-8-4-1
 \end{itemize}	
 
After each hidden layer in all the above networks, a dropout layer with probability as $0.4$ has been added to prevent overfitting. The hidden layers use ELU as the activation function, while the output layer uses sigmoid activation (since the output is a probability for binary classification). Adam optimizer with the learning rate of $0.001$ was used for training. Since it is a binary classification problem, the loss function used is binary cross-entropy. The differentiable nature of binary cross-entropy loss function will allow fast convergence. The entire dataset is split into $80$:$20$ where $80$\% is used for training, and the remaining $20$\% is used for testing. The training set is further split to give a 20\% validation split to analyze and tune the model learning during the training phase. The overall training is performed for $200$ epochs with a batch size of $32$. During training, the accuracy of the training and validation set was low during the initial epochs, but as the model starts to learn the accuracy increased in the future epochs. Detailed performance results of DNNs for malware detection are listed in Table \ref{big_table} and discussed in section \ref{discussion}.

\subsection{Discussion}\label{discussion}

\begin{table}[htbp]
	\centering
	\caption{Performance of different Classification Algorithms combined Feature Reduction}
	\label{big_table}
	\tiny
\begin{tabular}{|c|c|c|c|c|c|c|c|c|c|}
\hline
\textbf{\begin{tabular}[c]{@{}c@{}}Feature\\Reduction\end{tabular}} & \textbf{\begin{tabular}[c]{@{}c@{}}No of\\ features\end{tabular}} & \textbf{\begin{tabular}[c]{@{}c@{}}Classification\\Algorithm\end{tabular}} & \textbf{Accuracy} & \textbf{TPR}  & \textbf{TNR}  & \textbf{AUC}  & \textbf{F1 Score} & \textbf{\begin{tabular}[c]{@{}c@{}}Train Time\\ (sec)\end{tabular}} & \textbf{\begin{tabular}[c]{@{}c@{}}Test Time\\ (sec)\end{tabular}} \\ \hline
Original Data              & 256                                                               & DT\rlap{\textsuperscript{5}}                  & 93.0              & 94.7          & 91.4          & 93.1          & 93.0              & 6.65955                                                             & 0.01028                                                            \\
VT\rlap{\textsuperscript{1}}                         & 30                                                                & DT                  & 91.6              & 93.3          & 90.0          & 91.6          & 91.6              & 1.20814                                                             & 0.00363                                                            \\
PCA\rlap{\textsuperscript{2}}                        & 15                                                                & DT                  & 90.4              & 92.3          & 88.5          & 90.4          & 90.4              & 0.73475                                                             & \textbf{0.00268}                                                   \\
AE-1L\rlap{\textsuperscript{3}}                      & 64\rlap{\textsuperscript{*}}    & DT                  & 91.4              & 93.3          & 89.6          & 91.4          & 91.5              & 3.18144                                                             & 0.00551                                                            \\
AE-3L\rlap{\textsuperscript{4}}                      & 16*                                                               & DT                  & 90.5              & 92.7          & 88.4          & 90.5          & 90.6              & 3.10379                                                             & 0.00565                                                            \\ \hline
Original Data              & 256                                                               & kNN\rlap{\textsuperscript{6}}                 & 92.3              & 94.7          & 90.0          & 97.0          & 92.4              & 1.22648                                                             & 24.39308                                                           \\
VT                         & 30                                                                & kNN                 & 91.1              & 94.2          & 88.1          & 96.3          & 91.2              & 0.12977                                                             & 1.23292                                                            \\
PCA                        & 15                                                                & kNN                 & 90.5              & 93.7          & 87.5          & 95.9          & 90.7              & \textbf{0.05033}                                                    & 0.57151                                                            \\
AE-1L                      & 64*                                                               & kNN                 & 91.5              & 95.3          & 87.8          & 96.3          & 91.7              & 3.50612                                                             & 20.80623                                                           \\
AE-3L                      & 16*                                                               & kNN                 & 89.9              & 93.2          & 86.7          & 95.7          & 90.1              & 3.97619                                                             & 19.77041                                                           \\ \hline
Original Data              & 256                                                               & SVM\rlap{\textsuperscript{7}}                 & 84.5              & 96.3          & 73.1          & 93.3          & 86.0              & 108.1359                                                            & 18.91573                                                           \\
VT                         & 30                                                                & SVM                 & 79.6              & 98.2          & 61.5          & 88.6          & 82.6              & 16.00776                                                            & 2.14833                                                            \\
PCA                        & 15                                                                & SVM                 & 77.0              & \textbf{98.7} & 56.0          & 85.0          & 80.9              & 12.65536                                                            & 1.17977                                                            \\
AE-1L                      & 64*                                                               & SVM                 & 82.5              & 96.0          & 69.4          & 92.6          & 84.4              & 119.1237                                                            & 20.51435                                                           \\
AE-3L                      & 16*                                                               & SVM                 & 79.9              & 97.4          & 62.9          & 89.3          & 82.6              & 131.9112                                                            & 22.92997                                                           \\ \hline
Original Data              & 256                                                               & RF\rlap{\textsuperscript{8}}                  & \textbf{95.7}     & 96.4          & \textbf{95.1} & \textbf{99.4} & \textbf{95.7}     & 21.4354                                                             & 0.27449                                                            \\
VT                         & 30                                                                & RF                  & 94.5              & 96.0          & 94.3          & 99.1          & 94.5              & 14.54262                                                            & 0.25277                                                            \\
PCA                        & 15                                                                & RF                  & 93.2              & 94.9          & 91.5          & 98.7          & 93.2              & 12.49487                                                            & 0.2612                                                             \\
AE-1L                      & 64*                                                               & RF                  & 94.3              & 96.1          & 92.6          & 99.0          & 94.3              & 15.76513                                                            & 0.27906                                                            \\
AE-3L                      & 16*                                                               & RF                  & 93.0              & 95.1          & 90.9          & 98.6          & 93.0              & 15.30469                                                            & 0.2838                                                             \\ \hline
Original Data              & 256                                                               & AdaBoost\rlap{\textsuperscript{9}}            & 92.9              & 94.2          & 91.6          & 98.2          & 92.8              & 65.8992                                                             & 0.38113                                                            \\
VT                         & 30                                                                & AdaBoost            & 90.3              & 93.8          & 86.8          & 96.3          & 90.5              & 16.54785                                                            & 0.25632                                                            \\
PCA                        & 15                                                                & AdaBoost            & 88.0              & 92.6          & 83.5          & 95.0          & 88.4              & 11.99837                                                            & 0.2447                                                             \\
AE-1L                      & 64*                                                               & AdaBoost            & 90.1              & 93.0          & 87.4          & 96.7          & 90.3              & 53.41497                                                            & 0.37953                                                            \\
AE-3L                      & 16*                                                               & AdaBoost            & 87.6              & 91.9          & 83.4          & 95.2          & 87.9              & 52.36362                                                            & 0.37849                                                            \\ \hline
Original Data              & 256                                                               & DNN-2L\rlap{\textsuperscript{10}}              & 91.7              & 95.5          & 87.9          & 91.7          & 91.9              & 408.0807                                                            & 0.92617                                                            \\
VT                         & 30                                                                & DNN-2L              & 86.6              & 96.6          & 76.8          & 86.7          & 87.7              & 392.182                                                             & 1.49585                                                            \\
PCA                        & 15                                                                & DNN-2L              & 79.4              & 97.6          & 61.6          & 79.6          & 82.4              & 418.8092                                                            & 1.87103                                                            \\
AE-1L                      & 64*                                                               & DNN-2L              & 89.8              & 92.9          & 86.8          & 89.9          & 90.0              & 409.074                                                             & 1.06848                                                            \\
AE-3L                      & 16*                                                               & DNN-2L              & 87.1              & 93.5          & 80.8          & 87.2          & 87.7              & 421.9584                                                            & 1.21068                                                            \\ \hline
Original Data              & 256                                                               & DNN-4L\rlap{\textsuperscript{11}}              & 93.7              & 93.4          & 94.0          & 93.7          & 93.7              & 561.4316                                                            & 2.59779                                                            \\
VT                         & 30                                                                & DNN-4L              & 89.5              & 92.5          & 86.5          & 89.5          & 89.7              & 542.6468                                                            & 3.921                                                              \\
PCA                        & 15                                                                & DNN-4L              & 84.9              & 96.9          & 73.2          & 85.0          & 86.4              & 531.3042                                                            & 4.63093                                                            \\
AE-1L                      & 64*                                                               & DNN-4L              & 91.9              & 92.1          & 91.8          & 91.9          & 91.9              & 536.8712                                                            & 2.8647                                                             \\
AE-3L                      & 16*                                                               & DNN-4L              & 89.6              & 93.7          & 85.5          & 89.6          & 89.9              & 561.4489                                                            & 3.21776                                                            \\ \hline
Original Data              & 256                                                               & DNN-7L\rlap{\textsuperscript{12}}              & 91.8              & 87.5          & 92.6          & 91.7          & 91.3              & 663.8196                                                            & 5.68757                                                            \\
VT                         & 30                                                                & DNN-7L              & 91.7              & 90.4          & 93.0          & 91.7          & 91.5              & 648.3339                                                            & 8.08047                                                            \\
PCA                        & 15                                                                & DNN-7L              & 83.6              & 97.9          & 68.6          & 83.8          & 85.7              & 714.5312                                                            & 9.26997                                                            \\
AE-1L                      & 64*                                                               & DNN-7L              & 91.6              & 89.5          & 93.6          & 91.5          & 91.3              & 697.7748                                                            & 6.12618                                                            \\
AE-3L                      & 16*                                                               & DNN-7L              & 90.3              & 92.5          & 88.2          & 90.3          & 90.4              & 730.9962                                                            & 6.87906                                                            \\ \hline
\end{tabular}
	{\\\scriptsize\textsuperscript{1}Variance Threshold}
	{\scriptsize\textsuperscript{2}Principal Component Analysis}
	{\scriptsize\textsuperscript{3}1-layer Auto Encoder}
	{\scriptsize\textsuperscript{4}3-layer Auto Encoder}
	{\scriptsize\textsuperscript{5}Decision Tree}
	{\scriptsize\textsuperscript{6}k-Nearest Neighbour}
	{\scriptsize\textsuperscript{7}Support Vector Machine}
	{\scriptsize\textsuperscript{8}Random Forest}
	{\scriptsize\textsuperscript{9}Adaptive Boosting}
	{\scriptsize\textsuperscript{10}2-layer Deep Neural Network}
	{\scriptsize\textsuperscript{11}4-layer Deep Neural Network}
	{\scriptsize\textsuperscript{12}7-layer Deep Neural Network}
\end{table}

Table \ref{big_table} illustrates the performance of different Android malware detection models based on different evaluation metrics

In terms of Accuracy, RF outperforms all other machine learning and deep neural network models. It achieves the highest accuracy of $95.7\%$ when trained on original data. RF achieved the second \& third highest accuracy with VT \& AE-1L data reduction technique.

As far as AUC is concerned, RF with any data reduction technique is more balanced \& achieves more area under curve compared to any other classifier. RF (original data) attains AUC of $99.4$, followed by RF (VT data), RF (AE-1L data), \& RF (PCA data). High AUC score of RF models signifies that it can fit both malware and benign class properly without much variance.

Analyzing the recall, SVM achieves highest TPR of $98.7$ with PCA data followed by SVM (VT data), DNN-7L (PCA data) and DNN-2L (PCA data). Also for all the above models, higher TPR was achieved at the cost of very low TNR. In other words, the above models form a decision boundary which is skewed and highly favours the positive class: thus classifying even benign applications as malicious, which makes malware detection system unreliable. The exact number of false alarms by SVM (PCA data) and SVM (VT data) were $2517$ and $2203$ respectively, which is very high for any real-time deployment. On the other hand, RF (original data) produces the best TNR of $95.1$ followed by RF (VT data). A point to note here is RF (original data) archives less number of FP (280) and FN (201), making it more stable and reliable classifier. As expected SVM (PCA data) produces the worst TNR of $56.0$ since it is overfitting the malware class.

Model building (training time) is a one-time activity. Thus after building the classification model once, it can iteratively be used for testing new Android applications. kNN (PCA data) and kNN (VT data) take the least time to train model since kNN classifier does not build a model as such. kNN only performs pre-processing during model training, and thus it takes more time during the testing phase as it is seen for kNN (PCA data) and kNN (VT data). Also, the performance of kNN becomes worse as numbers of dimensions are increased to original data because of the Euclidian distance calculation to find neighbours. Also, while comparing the training time, all deep neural networks have a higher training time than any machine learning model, which can be attributed to the multiple epochs over which these networks need to be trained.

For the testing time, tree-based classifiers like DT (PCA data) and DT (VT data) perform the best. Also, RF (VT data) and RF (PCA data) take 0.25 sec and 0.26 sec respectively for testing with an ensemble of 100 trees. In the above cases, PCA data and VT data are performing better because they have less number of features as compared to original data. Thus it is a tradeoff between accuracy and testing time.

\subsection{Malware Detection Models based on classification integrated with clustering} \label{clustering}

As discussed in section \ref{discussion}, many classifiers (namely SVM (original data), SVM (VT data), DNN-2L (VT data) etc.) were overfitting the malware class (vector space) which decreases the overall performance of the Android malware detection system. Now we plan to divide the vector space ($11138$ $ \times $ $256$) with clustering into smaller vector spaces (clusters). The idea is instead of building a highly complex malware detection model which may underfit/overfit a particular class, the vector space itself can be segregated into smaller vector spaces (clusters). Then different/same detection models can be trained on each of these smaller vector spaces (clusters), thus solving overall underfitting/overfitting problem. The above approach is intuitive as well because malicious Android applications come from a very restricted vector space often bounded by their malware family (refer Fig \ref{fig/pca_dis}). At the same time, there is no such limitation with the benign applications which are very diverse (refer Fig \ref{fig/pca_dis}). Thus we chose five different clustering algorithms (k-means, Agglomerative, BIRCH, DBSCAN, and GMM). The first $3$ algorithms use a centre-based clustering approach while DBSCAN uses density measure, and GMM works on the probability distribution.

\begin{figure}[htbp]
	\centering
	\begin{minipage}[h]{0.49\textwidth}
    	\includegraphics[width=1\linewidth]{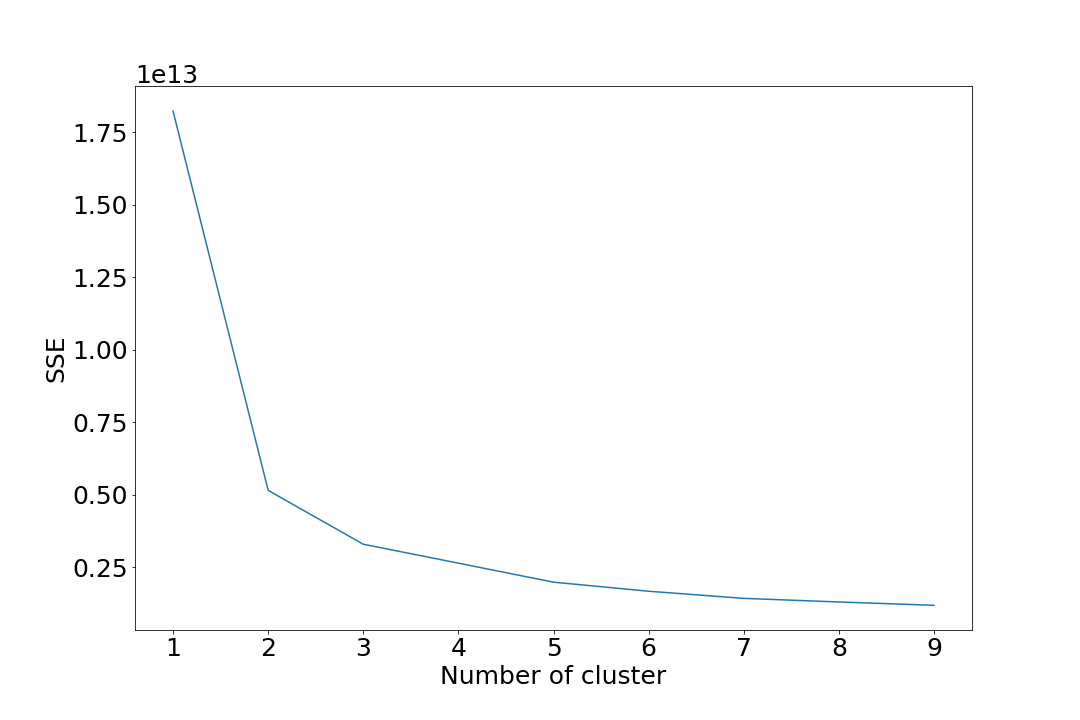}
    	\caption{Elbow Method (k-means Clustering)}
    	\label{fig/Elbow}
	\end{minipage}
	\hfill
	\begin{minipage}[h]{0.49\textwidth}
	\includegraphics[width=1\linewidth]{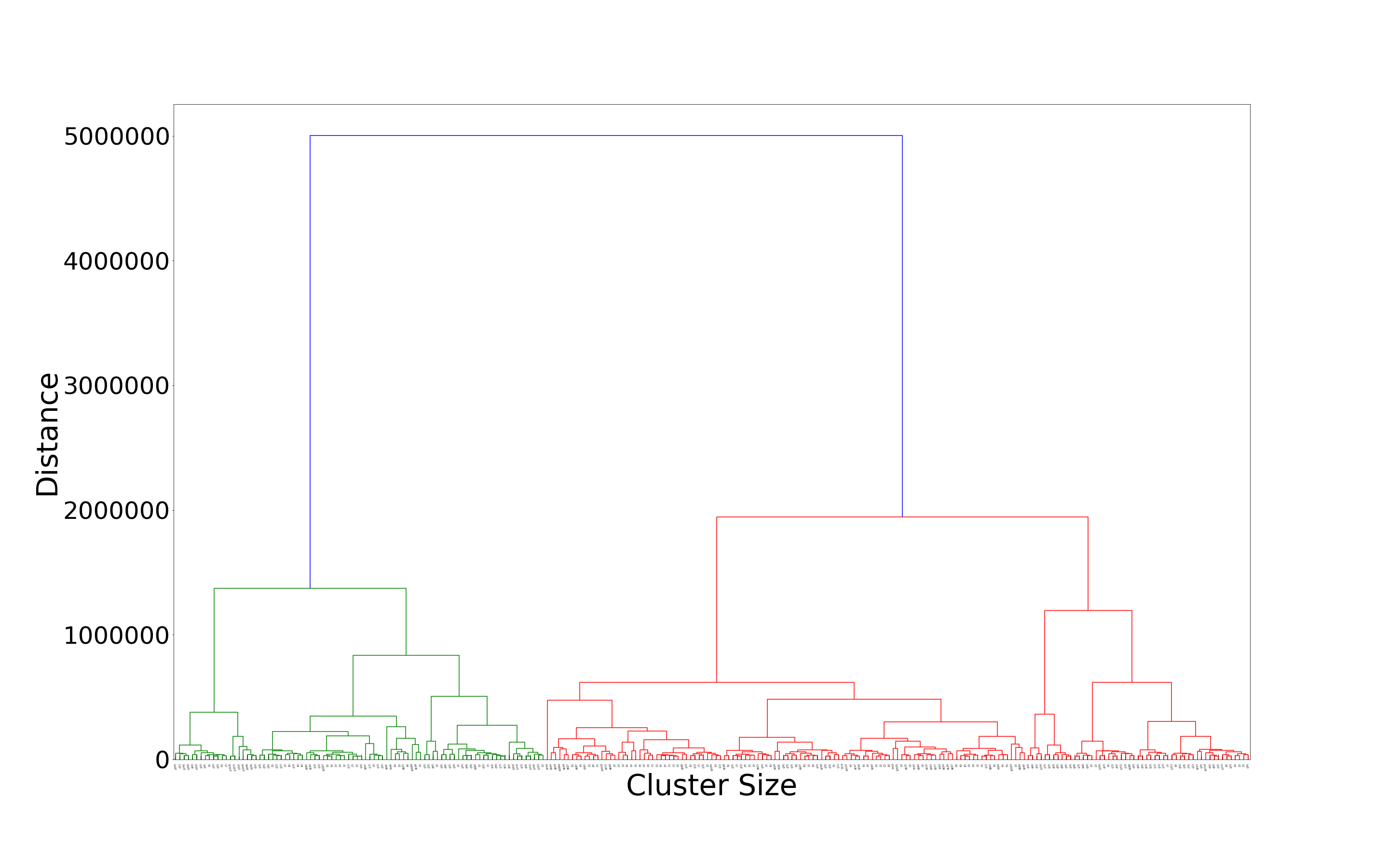}
	\caption{Dendogram (Agglomerative Clustering)}
	\label{fig/Dendogram}
	\end{minipage}
\end{figure}

\textbf{k-means} follows the centre-based clustering approach with a cost function set to minimize SSE. However, it cannot find an optimal number of cluster (k) on its own. Thus we first used k-means clustering on the dataset to form clusters and then used the elbow method, Silhouette Score and Calinksi-Harabaz Score to find the optimal number of clusters. The elbow method analysis (refer figure \ref{fig/Elbow}) suggests that the optimal number of cluster(s) in the dataset is either $2$ or $3$. Also, the result of Silhouette Score and Calinksi-Harabaz Score shown in Table \ref{cluster_table} concludes $k=2$ as an optimal value. The highest Silhouette Score ($0.74031$) and Calinksi-Harabaz Score ($27808.68$) among all the clustering algorithms was achieved by k-means clustering algorithm with (k=$2$). \textbf{Agglomerative clustering} with the bottom-up approach was used with Dendogram (refer figure \ref{fig/Dendogram}), Silhouette Score and Calinksi-Harabaz Score to find the optimal number of clusters. We performed cluster analysis with cluster numbers as $2$, $3$, $4$, $5$ and achieved highest Silhouette Score ($0.72839$) and Calinksi-Harabaz Score ($24025.74$) when the number of clusters was set as $2$ (refer Table \ref{cluster_table}). \textbf{BIRCH} is used to validate the scalability of the model, and it again follows the centre-based clustering approach. The highest Silhouette Score (0.72852) and Calinksi-Harabaz Score (24055.01) (refer Table \ref{cluster_table}) was yet again achieved at the number of clusters as $2$ during cluster analysis with different cluster number ($2$, $3$, $4$, $5$). \textbf{GMM and DBSCAN} performed poorly on the dataset with Silhouette Score and Calinksi Harabaz Score being the lowest among all the clustering algorithm at $0.20257$ and $8117.87$ with GMM (refer Table \ref{cluster_table}), and $0.55487$ and $1348.71$ with DBSCAN (refer Table \ref{cluster_table}) respectively.

%

% %
% \begin{table}[ht]
% 	\centering
% 	\abovecaptionskip=5pt
% 	\caption{Cluster formation using k-means clustering}
% 	\label{kmeans_table}
% 	\tiny
% 	\begin{tabular}{|c|c|c|}
% 		\hline
% 		\textbf{Number of Clusters} & \textbf{Silhouetee Score} & \textbf{Calinski Harabaz Score} \\ \hline
% 		2                           & 0.74031                   & 27808.68                        \\ \hline
% 		3                           & 0.70380                   & 24853.64                        \\ \hline
% 		4                           & 0.61534                   & 21623.26                        \\ \hline
% 		5                           & 0.61141                   & 22322.77                        \\ \hline
% 	\end{tabular}
% \end{table}
% %

% Please add the following required packages to your document preamble:
% \usepackage{multirow}
\begin{table}[htbp]
\centering
\caption{Cluster formation using different Clustering Algorithms}
\small
\label{cluster_table}
\begin{tabular}{|c|c|c|c|}
\hline
\textbf{\begin{tabular}[c]{@{}c@{}}Clustering\\ Algorithm\end{tabular}}             & \textbf{\begin{tabular}[c]{@{}c@{}}No of\\ Clusters\end{tabular}} & \textbf{\begin{tabular}[c]{@{}c@{}}Silhouetee\\ Score\end{tabular}} & \textbf{\begin{tabular}[c]{@{}c@{}}Calinski Harabaz\\ Score\end{tabular}} \\ \hline
\multirow{4}{*}{\begin{tabular}[c]{@{}c@{}}k-means\\ Clustering\end{tabular}}       & 2                                                                 & 0.74031                                                             & 27808.68                                                                  \\ \cline{2-4} 
                                                                                    & 3                                                                 & 0.70380                                                             & 24853.64                                                                  \\ \cline{2-4} 
                                                                                    & 4                                                                 & 0.61534                                                             & 21623.26                                                                  \\ \cline{2-4} 
                                                                                    & 5                                                                 & 0.61141                                                             & 22322.77                                                                  \\ \hline
\multirow{4}{*}{\begin{tabular}[c]{@{}c@{}}Agglomerative\\ Clustering\end{tabular}} & 2                                                                 & 0.72839                                                             & 24025.74                                                                  \\ \cline{2-4} 
                                                                                    & 3                                                                 & 0.70970                                                             & 20666.88                                                                  \\ \cline{2-4} 
                                                                                    & 4                                                                 & 0.55662                                                             & 18818.29                                                                  \\ \cline{2-4} 
                                                                                    & 5                                                                 & 0.55554                                                             & 19256.05                                                                  \\ \hline
\multirow{4}{*}{\begin{tabular}[c]{@{}c@{}}BIRCH\\ Clustering\end{tabular}}         & 2                                                                 & 0.72852                                                             & 24055.01                                                                  \\ \cline{2-4} 
                                                                                    & 3                                                                 & 0.70320                                                             & 21479.21                                                                  \\ \cline{2-4} 
                                                                                    & 4                                                                 & 0.59214                                                             & 20325.78                                                                  \\ \cline{2-4} 
                                                                                    & 5                                                                 & 0.58732                                                             & 19735.05                                                                  \\ \hline
\multirow{4}{*}{\begin{tabular}[c]{@{}c@{}}GMM\\ Clustering\end{tabular}}           & 2                                                                 & 0.20257                                                             & 8117.87                                                                   \\ \cline{2-4} 
                                                                                    & 3                                                                 & 0.20257                                                             & 2687.21                                                                   \\ \cline{2-4} 
                                                                                    & 4                                                                 & 0.11263                                                             & 1634.45                                                                   \\ \cline{2-4} 
                                                                                    & 5                                                                 & 0.25938                                                             & 5188.62                                                                   \\ \hline
\multirow{4}{*}{\begin{tabular}[c]{@{}c@{}}DBSCAN\\ Clustering\end{tabular}}        & 5000 (eps)                                                        & 0.55487                                                             & 1348.71                                                                   \\ \cline{2-4} 
                                                                                    & 10000 (eps)                                                       & 0.53745                                                             & 3381.47                                                                   \\ \cline{2-4} 
                                                                                    & 15000 (eps)                                                       & 0.67839                                                             & 3638.12                                                                   \\ \cline{2-4} 
                                                                                    & 20000 (eps)                                                       & 0.68565                                                             & 4719.16                                                                   \\ \hline
\end{tabular}
\end{table}

Finally, we chose the k-means clustering algorithm (with k=$2$) for cluster formation in the dataset since it has achieved the highest Silhouette Score and Calinksi-Harabaz Score among all the other algorithms (refer Table \ref{cluster_table}). After clustering the number of Android applications in Cluster-1 and Cluster-2 were $8,344$ and is $2,794$ respectively.

\textbf{Cluster-1 Analysis:} The Cluster-1 contains both malicious and benign applications. So we applied all the classifiers with similar parameters (Table \ref{big_table}) and achieved the results (refer to Table \ref{table_cluster_0}). Once again tree-based classifiers outperform all the classifiers. RF and DT have shown improvement in all metrics where RF outperforms DT. There has been $1.2$\% improvement in the accuracy and F1 score of RF from without the cluster approach. Other classifiers also performed better as compared to without the clustering approach.

\begin{table}[htbp]
	\centering
	\caption{Performance of Different Classifiers in Cluster-1}
	\label{table_cluster_0}
	\scriptsize

\begin{tabular}{|c|c|c|c|c|c|c|c|c|}
\hline
\textbf{\begin{tabular}[c]{@{}c@{}}Feature   \\ Reduction\end{tabular}} & \textbf{Classifier} & \textbf{Accuracy} & \textbf{TPR}  & \textbf{TNR}  & \textbf{AUC}  & \textbf{F1 Score} & \textbf{\begin{tabular}[c]{@{}c@{}}Train Time\\ (sec)\end{tabular}} & \textbf{\begin{tabular}[c]{@{}c@{}}Test Time\\ (sec)\end{tabular}} \\ \hline
Original Data                                                        & DT\rlap{\textsuperscript{1}}                  & 93.7              & 94.7          & 91.6          & 93.7          & 93.6              & 4.31299                                                             & \textbf{0.00936}                                                   \\
Original Data                                                        & kNN\rlap{\textsuperscript{2}}                 & 93.8              & 92.7          & 94.0          & 97.2          & 92.4              & \textbf{0.84515}                                                    & 18.99223                                                           \\
Original Data                                                        & SVM\rlap{\textsuperscript{3}}                 & 80.9              & 93.3          & 68.5          & 92.0          & 81.0              & 92.24813                                                            & 17.99223                                                           \\
Original Data                                                        & RF\rlap{\textsuperscript{4}}                  & \textbf{96.9}     & 96.2          & \textbf{97.9} & \textbf{99.6} & \textbf{96.8}     & 16.26439                                                            & 0.21686                                                            \\
Original Data                                                        & AdaBoost\rlap{\textsuperscript{5}}            & 92.9              & 93.3          & 92.5          & 97.4          & 92.3              & 57.96454                                                            & 0.25266                                                            \\
Original Data                                                        & DNN-2L\rlap{\textsuperscript{6}}              & 92.1              & 89.5          & 94.7          & 92.1          & 91.6              & 332.88325                                                           & 0.74821                                                            \\
Original Data                                                        & DNN-4L\rlap{\textsuperscript{7}}              & 94.1              & \textbf{96.3} & 89.9          & 93.1          & 93.2              & 454.42803                                                           & 1.97307                                                            \\
Original Data                                                        & DNN-7L\rlap{\textsuperscript{8}}              & 93.7              & 91.0          & 96.4          & 93.7          & 93.5              & 583.60624                                                           & 4.18910                                                            \\ \hline
\end{tabular}

	{\scriptsize\textsuperscript{1}Decision Tree}
	{\scriptsize\textsuperscript{2}k-Nearest Neighbour}
	{\scriptsize\textsuperscript{3}Support Vector Machine}
	{\scriptsize\textsuperscript{4}Random Forest}
	{\scriptsize\textsuperscript{5}Adaptive Boosting}
	{\scriptsize\textsuperscript{6}2-layer Deep Neural Network}
	{\scriptsize\textsuperscript{7}4-layer Deep Neural Network}
	{\scriptsize\textsuperscript{8}7-layer Deep Neural Network}

\end{table}

\textbf{Cluster-2 Analysis:} Interestingly Cluster-2 had over $98$\% of benign applications with only $2$\% of malicious samples. On further analysis of malware applications in the cluster, we found that the majority of them belong to only two families (\textit{Fatakr} and \textit{Steek}). Thus it clearly shows instead of building a single complex classifier for the complete dataset, segregation of vector space can give better insights on the data and yield better results.

\section{Conclusions and Future Work}

Today Android mobile phones are growing exponentially, but traditional malware detection systems are failing to cope up with the volume, velocity, and sophistication of malware attacks performed on these devices. In this paper, we proposed to used machine learning and deep neural network integrated with clustering for detection of malicious Android applications. We conducted a comprehensive analysis with different feature reduction, classification and clustering techniques to propose effective and efficient Android malware detection models.

Our baseline experimental results for Android malware detection models show that RF built without feature reduction achieved the highest ROC ($99.4$), accuracy ($95.7\%$), TNR ($95.1$), and F1 score ($95.7$). In fact, RF also performed better regarding the accuracy, ROC and F1 metric with any feature reduction method vis-\`a-vis other models with the same feature reduction method. Also, RF is more balanced (TPR $\sim$ TNR) and can fit both the malware and benign classes without much variance. Regarding training and testing time, tree-based classifiers viz. (RF) performed better than other classifiers, and is approximately $100$ and $500$ faster than DNN models respectively. Analyzing the other classifiers (viz. kNN, SVM), we found that they tend to over-fit malware class despite using cross-validation during model construction. A possible explanation is malware variants from a malware family tend to be similar to each other. Thus will be projected close to each other in the vector space while benign samples are well separated in the feature vector.

Our empirical results (clustering integrated with classification) show further improvement in AUC ($99.6$) and accuracy of RF ($96.9\%$) in the cluster-1 and also direct identification of malware applications of the two families in another cluster. The experimental results shows that constructing a single highly complex detection model might overfit/underfit data and suffer from the curse of dimensionality. Thus first segregation of vector space (clusters) using clustering followed by classification can further improve the performance of detection models.

Another contribution of our study is the performance of DNN for Android malware detection. We designed autoencoders (AE-1L (shallow) and AE-3L (deep)) and DNN of different sizes (DNN-2L, DNN-4L, and DNN-7L). Surprisingly none of the combinations of autoencoder and DNN performed well. One explanation could be a combination of complex feature reduction function, and sophisticated classifier leads to overfitting despite using drop out for generalization.  

Further, it will be interesting to see the performance of other deep learning based feature reduction techniques like sparse, denoising, variational based autoencoders  coupled with classification techniques like Hopfield Network, Sequence-to-sequence model etc.

\bibliographystyle{plain}
\bibliography{main}

\end{document}